\date{Version 3; January 21, 2011}
\documentclass[prl,aps,twocolumn,showpacs,superscriptaddress,nofootinbib]{revtex4-1} 
\usepackage{amsmath,times}
\usepackage{amssymb,times}
\usepackage{amsthm,times}
\usepackage{enumerate,times}

\usepackage{graphicx,pgf}

\makeatletter
\let\Im\undefined

\DeclareMathOperator{\Im}{Im \,}

\DeclareMathOperator{\E}{\mathbb{E}}

\newcommand{\T}{\mathbb{T}}

\def\Ev#1{{\mathbb E}\left(#1 \right)}

\def\be{\begin{equation}}
\def\ee{\end{equation}} 

\makeatother

\begin{document}

\title{ Extended States in a Lifshitz Tail Regime for     
                     Random Schr\"odinger Operators on Trees}

\author{Michael Aizenman} 
\affiliation{Departments of Physics and Mathematics, Princeton University,  Princeton NJ 08544, USA} 
\author{Simone Warzel}
\affiliation{Zentrum Mathematik, TU M\"unchen,  Boltzmannstrasse 3, 85747 Garching, Germany}  

\pacs{73.20.Jc,73.20.Fz}

\begin{abstract}
We resolve an existing question concerning the location of the mobility edge for operators with a hopping term and a random potential on the Bethe lattice.  The model has been among the earliest studied for Anderson localization, and it continues to attract attention because of analogies which have been suggested with localization issues for many particle systems.   We find that extended states appear through  disorder enabled resonances  
well beyond the energy band of the operator's hopping term.  For weak disorder this includes a Lifshitz tail regime of very low density of states.
\end{abstract} 

\keywords{Anderson localization, mobility edge, Bethe Lattice}
\maketitle

\noindent
{\it Introduction.--}   
Operators which combine a  hopping term and random potential on tree graphs are among the earlier studied models of Anderson localization~\cite{A,AAT,AT}.   Nevertheless, some questions about their phase diagram have persisted and remained open at both the rigorous and non-rigorous levels~\cite{MD}.   
The subject has recently attracted additional interest because of analogies which were drawn between systems of many particles and that  of one particle on a graph of very high degree, where loop effects may be unimportant~\cite{AGKL,BAA}.   

Our purpose here is to describe a new result, whose mathematical details will be spelled elsewhere,  concerning the nature of the spectrum in an intermediate regime, where the density of states is extremely low but where nevertheless known criteria for localization do not apply.   The main result is the existence 
of extended states, and absolutely continuous spectrum.   Aside from answering a long open question concerning the location of the mobility edge, the proof may be of added interest, as it introduces  a new mechanism for the formation of extended states which may reflect dynamics driven by disorder-facilitated resonant tunneling.      

More explicitly, we discuss the spectral properties of a Hamiltonian  operator of the  form
\be  \label{eq:H}
H_\lambda  \ = \ T + \lambda \, V   \, , 
\ee 
acting in the space of square-summable functions over a regular tree graph $\T$ of degree $(K+1)$ (i.e., where each sites has that many neighbors), with $T$ being the generator of  `hopping' transitions between neighboring sites  (i.e., $T_{x,y}=\delta_{|x-y|,1}$), $V$ being a random potential and $ \lambda \geq 0 $ serving as a disorder parameter. 
It is  assumed here that: 
\begin{itemize} 
\item[]   $\{V_x\}_{x\in \T}$ form independent identically distributed random variables, with a probability distribution whose density $\rho(V)$ is continuous and positive on the entire line, with  a finite number of humps, and  $ \E\left[|V_x|^\tau\right] < \infty $ for some $ \tau >0 $.  (Examples include the Gaussian and the Cauchy distributions, as well as any linear combinations.)  
\end{itemize} 

For ergodic potentials, a class which includes the present case, the spectrum of $H_\lambda$ is (almost surely) a nonrandom set, which in our case changes discontinuously:  at $\lambda = 0$ it is just 
\be \label{sigmaT}
\sigma(T) \ = \ [-2\sqrt{K},2\sqrt{K}] \, , 
\ee 
whereas for $\lambda > 0$ the spectrum of $H_\lambda$ is the entire  real line (as is the case for $V$).  
The phenomena we discuss here are related to the occurrence of energy regimes with different spectral and dynamical types whose locations are also nonrandom and which are separated by a so called \emph{mobility edge}.  In one spectral regime the operator $H_\lambda$ has only localized eigenstates  and in the other its spectrum is expressed in extended (generalized) eigenstates,   which enable conduction.   These spectral types indicate the nature of the evolution 
of wave packets with energies in the specified regimes, and of the conductance properties of  an electron gas in the corresponding one-particle approximation.    
 
 Simple examples of two different spectral types are offered by the two components of $H_\lambda$.
The operator $V$ has almost surely only \emph{pure-point} spectrum: it has a countable collection of proper eigenstates ($ \{ | u \rangle \}_{u \in \T } $) -- delta functions associated with the  sites  of $ \T$ -- and the collection of its  eigenvalues ($\{V_u\}_{u\in \T}$)   
forms a  dense set over the full line. 
In contrast, $T$ has continuous, in fact \emph{absolutely continuous}, spectrum: it has only generalized (non-square-summable) eigenfunctions, of continuously varying energies  (and infinite degeneracy).   

 Naively, one could expect that at least in  the perturbative regimes the spectrum of the sum~\eqref{eq:H}  would resemble that of the dominant term.   That, however, is not quite the case.  As is well known, in one dimension disorder has a nonperturbative effect: even at weak disorder it causes complete localization~\cite{MT,GMP}.  We now show that -- somewhat conversely -- on tree 
 graphs (other than 1D) 
  extended states, and absolutely continuous spectrum, emerge through resonances in regimes where at first sight one could expect localization to dominate.    

\medskip 

\begin{figure} 
\begin{center}
\includegraphics[width=0.47\textwidth]{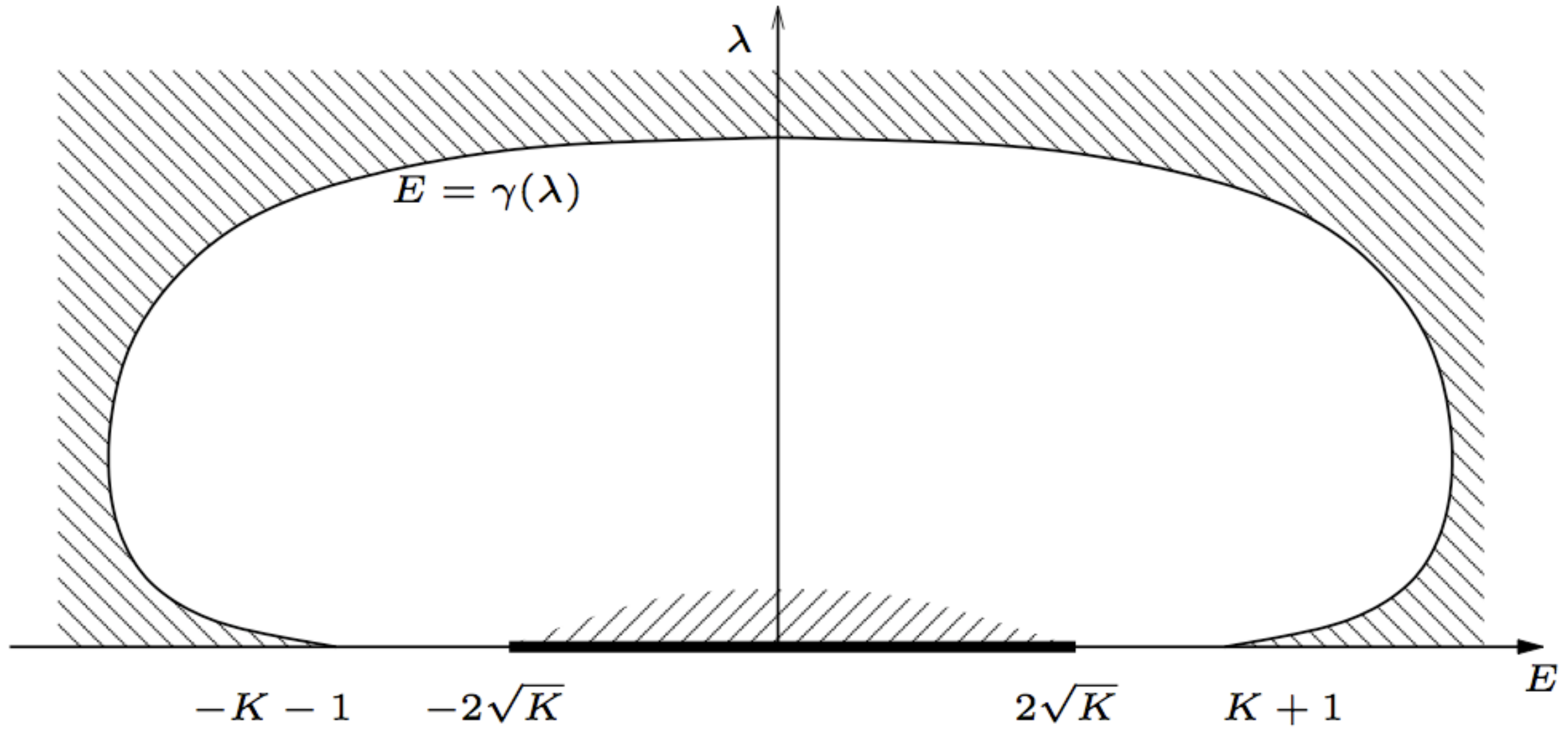}  
\caption{A sketch  of the previously known parts of the phase diagram.  The outer region is of proven localization, the smaller hatched region is of proven  delocalization.   The new result extends the latter up to the outer curve defined by $\varphi_\lambda(1;E) =- \log K  $, assuming equality holds only along a line.   
The depiction is schematic, however the points at which the curve meets the energy axis are stated exactly.
}\label{fig1} 
\end{center}
\end{figure}

\noindent{\it The puzzle left by past results.--} 
The phase diagram of  $H_\lambda$  was considered in the early works of Abou-Chacra, Anderson and Thouless~\cite{AAT,AT}.  Arguments and numerical work presented in~\cite{AT}  led the authors to surmise that a mobility edge exists at a location  which roughly corresponds to the outer curve in Fig.~\ref{fig1}, approaching limits close to $|E|=K+1$ as $\lambda \downarrow 0$. As noted there,  a puzzling aspect of this finding is that  this limit does not coincide with the spectral edge of $\sigma(T)$ (given by~\eqref{sigmaT}).  
The analysis offered in~\cite{AAT, AT}  mainly focused on the breakdown of a localization condition, without addressing the nature of the spectral regime beyond this stability edge.    Rigorous results on localization established the following~\cite{AM,A_wd}:
For energies in a regime of the form $|E|> \gamma(\lambda)$, with 
$ \gamma(\lambda)$  a function  satisfying $\lim_{\lambda \downarrow 0} \gamma(\lambda) = K+1$,  
as qualitatively depicted in Fig.~\ref{fig1}, 
with probability one the random operator exhibits spectral and dynamical localization.

Spectral localization means here that  in the specified range of energies the operator has only pure point spectrum, consisting of a dense set of non-degenerate proper eigenvalues whose eigenfunctions are exponentially localized.   Dynamical localization is expressed in the bound 
\begin{multline}  \label{eq:dynloc}
\sum_{|x|=R}\E \left(|\langle x| P_I(H_\lambda) \, e^{-itH_\lambda} |0\rangle |^2\right) \ 
   \le \ A_I \, e^{-\mu_\lambda(I) R} 
\end{multline}
for intervals $I$ lying  within the interior of the localization regime (the localization length $\mu_\lambda(I)^{-1}$ tends to zero as the  boundary of the localization regime is approached).  Here $P_I(H_\lambda)$ is the spectral projection onto the space spanned by states with energies in~$I$.

The curious gap between the edge of the proven localization regime (which is $|E|=K+1$) and the unperturbed spectrum ($|E|=2\sqrt{K}$)  
was considered in some detail in the study of Miller and Derrida~\cite{MD}.  It was noted there that for energies in the range $|E|> 2\sqrt{K}$
the mean density of states $ \rho_{\rm DOS}(E,\lambda) $ vanishes to all orders in $\lambda$, for  $\lambda \downarrow 0$.  E.g., 
in case of the   Gaussian distribution as $\rho_{\rm DOS}(E,\lambda) \approx \exp{\left(-C(E)/\lambda^2\right)}$.   Such rapid decay is characteristic of the so called \emph{Lifshitz tail} spectral regime, and in finite dimensions it is known to lead to localization.  
The conclusion of~\cite{MD}  was that the following question remained unresolved by the available methods.
\begin{description}
\item[Q] What is the nature of the spectrum for weak disorder  at 
 energies 
\be  \label{interE}
 2\sqrt{K} < |E| < K+1 \, 
 \ee 
(where the density of states vanishes faster than any power  of $\lambda$,  as $\lambda \downarrow 0$)?
\end{description}  
The rigorous results which were derived since then have only sharpened the question.  The existence of absolutely continuous spectrum, and diffusive dynamics~\cite{K}, has by now  also been established for the random operators discussed here.   However, till the present work this has been  accomplished only  for energies  $ |E|<2\sqrt{K}$, and  by arguments which address (in different ways~\cite{K,ASW,FHS}) regimes of small $\lambda$.  

Thus, the past results have covered two regimes whose boundaries, sketched in Figure~\ref{fig1}, do not connect.   The result presented here answers the question concerning the nature of the spectrum in the region between the two curves in Figure~\ref{fig1}.  To describe the proof, and link the new to the existing results, let us sketch some of the arguments which play a role in the derivation of localization.

%
%
\noindent {\it  Sketch of the localization analysis. --}  
Proofs of localization on tree graphs~\cite{AM,A_wd} have been  based on analysis of the fractional moments of the  Green function~\cite{multiscale}
 \be 
G_\lambda(0,x;z)  \ = \  \langle 0| (H_\lambda-z)^{-1} |x\rangle \, .
\ee 
Its relevance can be seen in the following  bound (which is valid for all graphs, not just trees) on the  transition probability from $ 0 $ to $ x $, for the time evolution generated by $H_\lambda $.   Restricting  to  energies in an interval 
$ I $ one has:
\begin{multline}  \label{eq:repFM}
\E \left(|\langle x| P_I(H_\lambda) \, e^{-itH_\lambda} |0\rangle |^2\right) \\
  \le \ C_{s,\lambda}
 \int_I   
\E\left( |G_\lambda(0,x;E+i0)|^s \right)\, dE 
\end{multline}
which holds (under some assumptions on the probability distribution of $V$) for any $s\in [0,1)$ at a constant $ C_{s,\lambda} <\infty$. 

Thus, a sufficient condition for dynamical localization in  $I$, in the sense of~\eqref{eq:dynloc}, is that for all $E\in I$: 
\be  \label{s-moment}
\sum_{ |x|=R} \Ev{|G_\lambda(0,x;E+i0)|^s} \,  \le\  e^{-\mu_\lambda(E) \, R} 
\ee
at some $s\in (0,1)$ and $\mu_\lambda(E) >0$.   Proofs of localization have proceeded by establishing this condition.

   To contrast the localization criterion with the new result, it is convenient to recast 
\eqref{s-moment} 
    in terms of the 
    function which we define  as~\cite{changelim}
\be  \label{eq:phi}
\varphi_\lambda(s;E)  \ = \  \lim_{|x|\to \infty} \frac{\log \mathbb{E}\left[\left|   G_\lambda(0,x;E+i0) \right|^s\right]}{|x|}  \, , 
\ee
 for $s\in [0,1)$, and for $s=1$:  $\varphi_\lambda(1;E) = \lim_{s\uparrow 1}\varphi_\lambda(s;E)$.

Clearly, on tree graphs 
 the localization condition \eqref{s-moment} holds wherever
\be  \label{phi<}
\varphi_\lambda(1;E) \ < \ - \log K  \, .  
\ee 
Our new result is that the opposite inequality implies  absolutely continuous spectrum.

\medskip 
 \noindent {\it The new result.--}    
The condition for absolutely continuous ($ac$)  spectrum  
throughout an energy interval $I$, is 
\be  \label{cond:ac}
\Im G_\lambda(0,0;E+i0) \ > \ 0  \quad \mbox{for almost all $E\in I$} \, .  
\ee 
For the random operator discussed here, if \eqref{cond:ac} holds with positive probability then it holds with probability one.   
One may note that    
$\pi^{-1}\Im G_\lambda(0,0;E+i0)$ is the density of the absolutely continuous component of the spectral measure associated with the state 
$| 0 \rangle $.   
The above condition  carries also direct  consequences for the dynamics: \\
\indent  {\it i.\/}~As explained in~\cite{MD,ASWb}, 
when particles of energy $E$ are send coherently down a wire which is connected to the point $x\in \T$,   the  reflection coefficient is less than $1$ and some of the current is  transmitted through the graph if and only if 
$\Im G_\lambda(x,x;E+i0) \ > \ 0 $.       \\  
\indent {\it ii.\/}~By the Riemann-Lebesgue lemma,  any state whose eigenfunction decomposition consists solely of states in the $ac$ spectrum  of $H_\lambda$ (a subspace on which the projection is denoted by $P_{ac}(H_\lambda)$) would,  in the course of the dynamics  generated by $H_\lambda$,  asymptotically  leave any finite region.  I.e., 
$$ \displaystyle  \lim_{t\to \infty} \sum_{|x| \leq R} |\langle x |  e^{-itH_\lambda} P_{ac}(H_\lambda)| \varphi \rangle |^2 = 0 \quad \mbox{ for any $ R > 0$.} $$

\medskip 
\noindent{\bf Theorem} {{\bf (Delocalization regime)}  \it
For the random operators considered here, \eqref{cond:ac} is satisfied, and hence the operators have almost surely absolutely continuous spectrum, throughout any interval where  
\begin{equation} \label{phi_ac}
\varphi_\lambda(1;E) \ > \ - \log K\, .
\end{equation}
Furthermore, the corresponding region in the $E\times \lambda$ `phase diagram'  includes for each energy $|E|<K+1$ an interval with a positive range of $\lambda>0$.   }
\smallskip

By  convexity, a sufficient condition for \eqref{phi_ac}  
is that the  Lyapunov exponent, $\displaystyle L_\lambda(E) = - \frac{\partial \varphi_\lambda }{\partial s}(0;E) $,  satisfies
\be  \label{lyapcond}
L_\lambda(E)   \  < \   \log K \, . 
\ee    
This yields a somewhat more tractable sufficiency criterion for delocalization, which is quite effective at weak disorder and at  high $K$.  The Lyapunov exponent yields  the Green function's typical rate of decay:  $|G(0,x;E+i0)| \approx e^{-L_\lambda(E) |x|}$, and it can be expressed as 
$ L_\lambda(E)  = -\E(\log |\Gamma_\lambda(0;E+i0) |) $, in terms of the quantity $\Gamma_\lambda(0;E+i0)$ which is defined in \eqref{eq:recur} below.    
It  is calculable for the Cauchy random potential, in which case
\eqref{lyapcond} holds at  $E=0$  for (exactly) $|\lambda| < K-1$.

At first sight, the  characterization of the mobility edge by $\varphi_\lambda(1;E) \ =  \ - \log K $  may be surprising, since  this has the appearance of a threshold for $\ell^1$  summability whereas  the transition from pure point spectrum corresponds to the loss of {\it square} summability, i.e., an $\ell^2$ condition.   On the tree, or any graph of exponential growth, the latter may continue to be satisfied well beyond the threshold for $\ell^1$  summability.    However, the criterion incorporates an important fluctuation effect.     Some light may be shed on  this phenomenon by the following heuristic argument~\cite{KSnote}.

\medskip 
\noindent{\it A heuristic explanation.--}  Let us first indicate why the operator has no pure-point  spectrum in any interval throughout which~\eqref{phi_ac} holds.  
The hallmark of such spectrum in a given interval~$I$, for an operator with random potential   is the almost-sure square summability of the Green function, at fixed $E\in I$:  $\sum_y |G_\lambda(x,y;E+i0)|^2 < \infty$, from any given site~$x$.  This is due to the {\it spectral averaging principle}, which implies that in the presence of disorder, such typical properties of the Green function are shared by the operator's eigenfunctions (cf.~\cite{SimWolff,A_wd}).  
Now, for energies $E$ in the localization regime 
\be \label{eigenexp}
G_\lambda(0,x;E+i0) \approx \sum_{n} \frac{1}{E_n-E} \, \varPsi_n(0)\, \overline{\varPsi}_n(x)  \, ,   
\ee 
where $\approx$ is used since we put aside the fact that the spectrum may include more distant, and hence non-singular terms where an integral expression is to be used.   
The localized eigenfunctions decay exponentially, each from its particular center, $x_n$, at an exponential rate which is typically given by the Lyapunov exponent.   For a given $R$ there will typically be about $\Delta E \, \rho_{\rm DOS}(E,\lambda) \,  K^R$  eigenfunctions localized in the vicinity of the shell $|x|= R$, with energies scattered within $E\pm \Delta E$.  The typical value of the smallest energy gap in such case would be 
\be 
\min_{n\, : |x_n|\approx R}  \{  |E_n-E| \}  \ \approx \  
\left[\rho_{\rm DOS}(E,\lambda) \,  K^R \right]^{-1}  \, . 
\ee 
For $x_{\min}$ -- the site where the minimum is attained --   the contribution of just the corresponding eigenfunction to the Green function would  typically be of the order of $ K^R \  e^{-R L_\lambda(E)}$. Making the seemingly reasonable assumption  that this relatively large term would not be exactly cancelled by the other contributions in \eqref{eigenexp}, one is led to expect that under the Lyapunov exponent condition \eqref{lyapcond}, if $E$ is in the pure point spectral regime then typically
\be   \label{maxG}
\max_{x\, : |x|=R} |G_\lambda(0,x;E+i0)| \ \gg  \  e^{\delta R}
\ee 
at some $\delta >0$.   As explained above, this contradicts the assumption of pure point spectrum.  

Taking into the account the contribution of eigenfunctions with a slower decay, through large deviation analysis, one arrives at the conclusion that \eqref{maxG} is to be expected also under the less stringent condition \eqref{phi_ac}, as stated in the theorem.   

The above argument is obviously incomplete, and at best it indicates only the existence of continuous spectrum, though not necessarily \emph{absolutely continuous} spectrum.   
The theorem states more: with probability one  $\Im G_\lambda(0,0;E+i0) > 0$.   
Following is a sketch of the proof of this assertion.

\medskip 
  The analysis on trees simplifies due to two relations: \\  
\emph{i.}  Upon the removal of the root of a rooted regular tree, the graph splits into a collection of $K$ trees which are similar to the original one, and whose roots form the set $\mathcal N^+_x$ of  the forward neighbors of $x$.  This leads to the \emph{recursion relation} (the `self-consistency' condition  of~\cite{AAT})
\be  \label{eq:recur}
\Gamma_\lambda(x;\zeta) = \Big( \lambda \, V_x - \zeta - \sum_{y \in \mathcal N^+_x} \Gamma_\lambda(y;\zeta) \Big)^{-1}  
\ee
with $  
\Gamma_\lambda(x;\zeta) :=   \langle  x |\,  (H_{\lambda}^{\mathbb{T}_x} -\zeta)^{-1} \, |x  \rangle \;    
$
the resolvent, at the extended complex energy parameter  $\zeta = E+i\eta$,   
of the operator obtained by restricting $H_\lambda$ to the subtree ($\mathbb{T}_x $) of  sites  which are beyond~$ x$ relative to the root.

\emph{ii.}  The other simplifying feature of trees is that the Green function $ G_\lambda(0,x;\zeta)$  factorizes into a product taken along the path ($0 \preceq u \preceq x$) from  the root to $x$: 
\be \label{factorization}
G_\lambda(0,x;\zeta) =  \langle 0 | \,  (H_\lambda-\zeta)^{-1} \, | x  \rangle = \prod_{0 \preceq u \preceq x } \Gamma_\lambda(u;\zeta)  \, .   
\ee  

By the above relations, for any integer $R\ge 0$: 
\be \label{55}
\Im \Gamma_\lambda(0;\zeta) \ \ge  \ \sum_{|x_+|=R} |G_\lambda(0,x;\zeta)|^2  \, \Im \Gamma(x_+;\zeta) 
\ee  
where $x$ is the site just below $x_+$ relative to the root. Equality holds  in case $\Im \zeta =0$ and the above relation is physically interpreted as the current conservation when feeding current through a wire at $ 0 $ (cf.~\cite{MD}).
To prove that the imaginary part does not vanish in the limit $ \Im \zeta \downarrow 0 $, we show the instability of the set of real distributions for $\Gamma_\lambda(0;\zeta)$ under the iterative relation \eqref{55}.   Key role is played by a large deviation analysis which is enabled by the above multiplicative structure.


\medskip
\noindent{\it Discussion.--}   Our analysis indicates that the relevant phenomenon for the transition from localization to continuous spectrum is  the formation of extended states through rare resonances between distant localization centers, which can be found due to  the exponential increase of surface at radius $R$.   This mechanism   does not apply on graphs of finite dimension.   However, we expect it to be of relevance also for  other hyperbolic graphs, including with loops, and possibly for many particle systems. 
It will also be interesting to better understand the implication of the picture 
presented above on the dynamics.   Analysis of the Green function suggests evolution by tunneling through forbidden regions.  The spread of an initially localized approximate eigenfunction may be rather non-uniform on scales smaller than  a  `tunneling distance', which in this regime is large compared to the lattice spacing.  
On larger scales 
(in both distance and time) the spread of the wave packet may become more uniform and asymptotically ballistic.    This appears different  than the more steady evolution which one may expect in the perturbative regime of $E\in \sigma(T)$ and  $\lambda$ 
small~\cite{K}. 
The transition between  the two  regimes of delocalization may be where the `tunneling distance' is comparable to the lattice spacing, and it may occur as a gradual crossover  without a sharply defined spectral edge.   
However, the rigorous analysis does not yet address these questions. 
Finally, let us note that the phase diagram  will be different in case of a uniformly bounded potential, such as the Anderson model, for which recent numerical results are presented in \cite{BST10}.  

We thank the  Departments of Physics and Mathematics at the  Weizmann Institute of Science for hospitality. 
This research was supported  by NSF grants DMS-0602360 (MA) and DMS-0701181 (SW), the A.~P.~Sloan Foundation~(SW), and BSF  710021. 


\begin{thebibliography}{99}

\bibitem{A} P.W. Anderson, Phys. Rev. {\bf 109},  1492 (1958). 

\bibitem{AAT}  R.~{Abou-Chacra}, P.~W. Anderson, D.~J. Thouless,
  J. Phys. C: Solid State Phys.  {\bf 6}, 1734 (1973).

\bibitem{AT}  R.~Abou-Chacra, D.~J. Thouless, 
 J. Phys. C: Solid State Phys. {\bf 7}, 65 (1974).
 
 \bibitem{MD}
J.~D. Miller, B.~Derrida,
\newblock 
J. Stat. Phys, {\bf 75}, 357 (1994).


\bibitem{AGKL}  B.L. Altshuler, Y. Gefen, A. Kamenev, L.S. Levitov,  
Phys. Rev. Lett. {\bf 78}, 2803  (1997).

\bibitem{BAA}
D.~M.~Basko, I.~L.~Aleiner, B.~L.~Altshuler, 
Annals of Physics {\bf 321}, 1126  (2006).


\bibitem{MT} N. F. Mott, W. D. Twose, Adv. Phys. {\bf 10}, 107 (1961).

\bibitem{GMP}   I. Goldsheid, S. Molchanov, L. Pastur, Funct. Anal. Appl. {\bf 11}, 1  (1977).    

\bibitem{A_wd}
M.~Aizenman, 
 Rev. Math. Phys. {\bf 6}, 1163 (1994).
 

\bibitem{AM}
M.~Aizenman, S.~Molchanov,
 Comm. Math. Phys. {\bf 157}, 245 (1993). 
 
 \bibitem{K}  A.~Klein, 
 Commun. Math. Phys., {177}, 755 (1996). Adv. Math.,
133, 163 (1998).
 
\bibitem{ASW}   
M.~Aizenman, R.~Sims, S.~Warzel,  
 Prob. Theor. Rel. Fields, {\bf 136}, 363  (2006). 

\bibitem{FHS} R. Froese, D. Hasler, W. Spitzer,
 Comm. Math. Phys. {\bf  269},  239--257  (2007).
 
 \bibitem{multiscale}
The  `multiscale method'~\cite{MultiScale} for a proof of localization could not be applied to trees because of their volume's  exponential growth.

 \bibitem{MultiScale}
 J. Fr\"ohlich, T. Spencer,
Comm. Math. Phys. {\bf 88}, 151 (1983).

\bibitem{changelim}
For $s<1$    the limits $\eta\downarrow0$ and $|x|\to \infty$  are  interchangeable, and hence $\varphi_\lambda(s;E) $  is decreasing in $s$.  This  is not the case  for $s\ge1$, where the expected value in~\eqref{eq:phi} diverges if  $\eta\downarrow 0$ is taken first, for  $E$ in the localization regime.


\bibitem{ASWb} M. Aizenman, R. Sims, S. Warzel, in: 
{\it Quantum Graphs and Their Applications}
(Snowbird, UT, 2005). 
Contemp. Math. {\bf 415}, AMS (2006).

\bibitem{KSnote}
The formulation of the mobility edge in terms of an $\ell^1$ condition appeared early on in~\cite{KS}.   However, that announcement was not followed by a proof.  So far the outlined method  found its expression only in the analysis of~\cite{K} which addresses small $\lambda$ and $E\in \sigma(T)$.

\bibitem{KS} H. Kunz, B. Souillard, J. Physique Lett. {\bf  44}, 411  (1983).


\bibitem{SimWolff}  B. Simon, T. Wolff, 
Comm. Pure Appl. Math. {\bf 39}, 75 (1986). 


\bibitem{BST10} G. Biroli, G. Semerjian, M. Tarzia,  Prog. Theor. Phys. Suppl. {\bf 184},  187 (2010).
\end{thebibliography}
\end{document}